\documentclass[sigplan,screen]{acmart}

\AtBeginDocument{%
  }

\setcopyright{acmlicensed}
\copyrightyear{2024}
\acmYear{2024}
\acmDOI{XXXXXXX.XXXXXXX}

\acmConference[]{}{}{}
\acmISBN{XXXXXXXXXXXX}

\usepackage{enumerate}
\usepackage{enumitem}
\usepackage{multirow}
\usepackage{soul}
\usepackage{color, xcolor}
\usepackage{acronym}
\usepackage{amsmath}
\usepackage{graphicx}

\begin{document}

\acrodef{GNNs}{graph neural networks}
\acrodef{SR}{Sequential recommendation}
\acrodef{RNNs}{recurrent neural networks}
\acrodef{CNNs}{convolutional neural networks}
\acrodef{SSL}{self-supervised learning}
\acrodef{MCs}{Markov chains}
\acrodef{GRUs}{gated recurrent units}
\acrodef{FFTs}{Fast Fourier transforms}
\acrodef{DL}{Deep Learning}
\acrodef{MLP}{Multilayer Perceptron}
\acrodef{CL}{contrastive learning}


\title{Learnable Sequence Augmenter for Triplet Contrastive Learning in Sequential Recommendation }


\author{WEI WANG}
\email{202120421@mail.sdu.edu.cn}
\orcid{0000-0002-7080-3381}
\affiliation{%
  \institution{School of Information Science and Engineering, Shandong University}
  \streetaddress{72 Binhai Rd}
  \city{QingDao}
  \country{CHINA}
  \postcode{266237}
}

\author{Yujie Lin}
\email{yu.jie.lin@outlook.com}
\orcid{0000-0002-2146-0626}
\affiliation{%
  \institution{Zhejiang Lab}
  \streetaddress{Kechuang Avenue, Zhongtai Sub-District}
  \city{Hangzhou}
  \country{CHINA}
  \postcode{311121}
}

\author{Pengjie Ren}
\email{renpengjie@sdu.edu.cn}
\orcid{0000-0003-2964-6422}
\affiliation{%
  \institution{School of Computer Science and Technology, Shandong University}
  \streetaddress{72 Binhai Rd}
  \city{QingDao}
  \country{CHINA}
  \postcode{266237}
}

\author{Zhumin Chen}
\email{chenzhumin@sdu.edu.cn}
\orcid{0000-0003-4592-4074}
\affiliation{%
  \institution{School of Computer Science and Technology, Shandong University}
  \streetaddress{72 Binhai Rd}
  \city{QingDao}
  \country{CHINA}
  \postcode{266237}
}

\author{Jianli Zhao}
\email{jlzhao@sdust.edu.cn}
\orcid{0000-0002-7291-9003}
\affiliation{%
  \institution{School of Computer Science and Engineering, Shandong University of Science and Technology}
  \streetaddress{579 Qianwangang Rd}
  \city{QingDao}
  \country{CHINA}
  \postcode{266590}
}

\author{Moyan Zhang}
\email{zmy20001122@163.com}
\orcid{0000-0001-6130-1286}
\affiliation{%
  \institution{School of Information Science and Engineering, Shandong University}
  \streetaddress{72 Binhai Rd}
  \city{QingDao}
  \country{CHINA}
  \postcode{266237}
}

\author{Xianye Ben}
\email{benxianye@gmail.com}
\orcid{0000-0001-8083-3501}
\affiliation{%
  \institution{School of Information Science and Engineering, Shandong University}
  \streetaddress{72 Binhai Rd}
  \city{QingDao}
  \country{CHINA}
  \postcode{266237, Corresponding author}
}

\author{Yujun Li}
\email{liyujun@sdu.edu.cn}
\orcid{0000-0003-4455-5991}
\affiliation{%
  \institution{School of Information Science and Engineering, Shandong University}
  \streetaddress{72 Binhai Rd}
  \city{QingDao}
  \country{CHINA}
  \postcode{266237, Corresponding author}
}

\renewcommand{\shortauthors}{WANG et al.}

\begin{abstract}
Most existing contrastive learning-based sequential recommendation (SR) methods rely on random operations (e.g., crop, reorder, and substitute) to generate augmented sequences. These methods often struggle to create positive sample pairs that closely resemble the representations of the raw sequences, potentially disrupting item correlations by deleting key items or introducing noisy iterac, which misguides the contrastive learning process.

To address this limitation, we propose \textbf{L}earnable sequence \textbf{A}ugmentor for triplet \textbf{C}ontrastive \textbf{L}earning in sequential \textbf{Rec}ommendation (LACLRec). Specifically, the self-supervised learning-based augmenter can automatically delete noisy items from sequences and insert new items that better capture item transition patterns, generating a higher-quality augmented sequence. Subsequently, we randomly generate another augmented sequence and design a ranking-based triplet contrastive loss to differentiate the similarities between the raw sequence, the augmented sequence from augmenter, and the randomly augmented sequence, providing more fine-grained contrastive signals. Extensive experiments on three real-world datasets demonstrate that both the sequence augmenter and the triplet contrast contribute to improving recommendation accuracy. LACLRec significantly outperforms the baseline model CL4SRec, and demonstrates superior performance compared to several state-of-the-art sequential recommendation algorithms.
\end{abstract}

\ccsdesc[500]{Information systems~Recommender systems}

\keywords{Sequential recommendation, Contrastive learning, Self-supervised learning}

\maketitle

\section{INTRODUCTION}
\acf{SR} \cite{1-,2-,3-,4-,5-}, which predicts user preferences based on their historical interactions, has garnered increased attention in recent recommendations studies. SR algorithms account for the temporal and sequential order of user behavior, capturing the transfer relationship between items, so as to recommend the next item for users according to the ordered interaction records over a period of time. 
Recently, \acf{SSL} has been introduced to sequential recommendation for extracting robust item correlations by semi-automatically exploiting raw item sequences \cite{15-ma,16-xia, 18-yu,30-}. Despite their effectiveness, sequential recommendation systems face notable challenges, including data sparsity and cold-start.

As an important branch of \ac{SSL}, \acf{CL} has been introduced into recommendation systems \cite{21-,22-,23-,24-} in recent years. Contrastive learning in recommendation systems mainly consists of two parts: constructing augmented data from the original user interaction data for training, and designing contrastive loss to provide additional supervisory signals \cite{25-}. The specific approach maximizes the similarity between positive pairs (augmented data from the same source) while improving discrimination ability to the negatives samples. Contrastive learning effectively alleviates the data sparsity and cold-start problems in recommendation systems, and \ac{CL}-based models \cite{26-,27-,28-,29-} have demonstrated more competitive recommendation performance.
\vspace{-0.5cm}
\begin{figure}[h]
  \setlength{\abovecaptionskip}{-0.1cm}
  \setlength{\belowcaptionskip}{-0.3cm}
  \centering
  \includegraphics[width=0.95\linewidth, height=6cm]{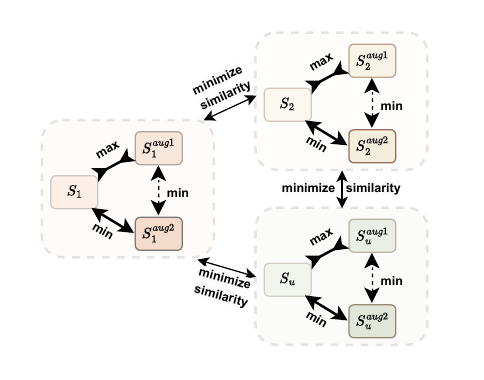}
  \caption{Illustration of triplet contrastive learning.}
  \Description{The figure is used to help readers understand what is presented in the introduction.}
  \label{fig1}
\end{figure}

However, existing \ac{CL}-based sequential recommendation methods still have the following shortcomings: (1) Most models such as CL4SRec \cite{17-xie} adopt random data augmentation to generate contrastive sequences, where mask and crop operations may delete key items and further amplify the data sparsity problem, and the reorder operation significantly disrupts the item transition correlations in the raw sequence. Moreover, the two contrastive sequences generated by the random method may be overly similar. (2) Although other works such as CoSeRec \cite{31-} and TiCoSeRec\cite{32-} adopt \ac{SSL} to perform substitute and insertion operations, they randomly select items to modify from the raw sequence, which still carries a high probability of deleting key items. (3) Real-world sequences may contain some noisy interactions (such as promotions), leading to negative feedback. Existing augmentation methods are not effective in deleting such noise. (4) Finally, the above works treat the two augmented sequences equally, and the raw sequence is not directly considered in the contrastive loss, leading to a lack of more fine-grained supervisory signals.

To address the aforementioned issues, we propose a triplet contrastive learning framework for sequential recommendation with a learnable sequence augmenter (LACLRec). Unlike existing methods, we utilize a trained \ac{SSL}-augmenter to generate augmented sequence from the raw sequence (referred to as \ac{SSL}-augmented sequence) and consider the differences in representation similarity among the \ac{SSL}-augmented sequence, random augmented sequence, and the raw sequence. Specifically, (1) We first disrupt the raw sequence by randomly deleting and inserting items, and the augmenter is trained to restore the disrupted sequence. This trained augmenter can automatically delete noise and insert new items, generating \ac{SSL}-augmented sequence without manually designed heuristics. (2) Then, we take one \ac{SSL}-augmented sequence and one random augmented sequence as positive pair, while using augmented sequences from other raw sequences as negative samples. By optimizing the contrastive loss $L_{cl}$, self-supervised signals are provided to the recommender. (3) Compared to random augmented sequence, we find that \ac{SSL}-augmented sequence enables the recommender to more accurately capture item transition correlations and predict the next interaction. Therefore, we design a triplet consisting of the raw sequence and its two augmented sequences, and introduce a BPR contrastive loss $L_{tri}$ to maximize the similarity between the raw sequence and the \ac{SSL}-augmented sequence, while minimizing the similarity between the raw sequence and the random augmented sequence. As shown in Figure \ref{fig1}, we aim to maximize the representational distance between different triplets, while within the triplet, the \ac{SSL}-augmented sequence provides more positive feedback. Finally, we jointly optimize $L_{cl}$ and $L_{tri}$, providing finer-grained contrastive signals for model training. We conduct extensive experiments on three datasets, and LACLRec significantly outperforms the baseline model CL4SRec, with a single metric improving by up to 13.5\%, demonstrating the effectiveness of the sequence augmenter and the triplet contrastive learning. LACLRec also outperforms several other state-of-the-art sequential recommendation models. 

To sum up, the main contributions of this work are as follows:
\begin{itemize}
    \item We propose a novel contrastive learning-based sequential recommendation framework (LACLRec), which incorporates a SSL sequence augmenter.
	\item The SSL-augmenter can automatically delete noise from raw sequence and insert new items, generating higher-quality augmented sequence.
    \item A ranking-based contrastive loss is designed to differentiate the representation similarities between the raw sequence, the randomly augmented sequence, and the SSL-augmented sequence, providing more fine-grained contrastive signals.
    \item We conduct extensive experiments to demonstrate the state-of-the-art performance of LACLRec. To facilitate reproducibility, we release the code and data at https://github.com/.
\end{itemize}

\section{RELATED WORK}
This section reviews \acf{CL}-based recommendation systems, categorizing them into two types: research on data augmentation methods and research on contrastive learning-based recommendation framework.

\subsection{Data Augmentation Research}
The main objective of contrastive learning is to maximize the similarity between positive samples while minimizing the similarity between positive and negative samples. Therefore, sampling positive and negative samples is a key challenge in contrastive learning. Many works have focused on developing better data augmentation methods to improve the performance of contrastive learning in recommendation systems.

\citet{17-xie} propose CL4SRec, an early work that applies contrastive learning to sequential recommendation. They design random augmentation methods such as mask, crop, and reorder, which enhance the accuracy of sequential recommendations. \citet{31-} introduce CoSeRec, which selects new items with the highest representation similarity and inserts or replaces them in the sequence at a predetermined ratio. \citet{32-} advocate for sequence augmentation from a time interval perspective and improve upon CoSeRec by generating augmented sequences with uniform time intervals. \citet{33-} propose a dual contrastive learning model that enhances both low-level (item) and high-level (item attribute) preference learning for users. \citet{34-} design rule-based augmentation, replacing items in the sequence based on item attributes. \citet{35-} introduce equivariant contrastive learning, which makes user representations sensitive to intrusive augmentations (such as substitute) and robust to mild augmentations (such as mask). \citet{36-} introduce spatiotemporal frequency domain techniques from computer vision and design a data augmentation strategy to model users' interest trends. \citet{37-} and \citet{38-} design intent contrastive learning frameworks, while \citet{39-} segment the original sequence into multiple subsequences, assuming that different subsequences with the same target item represent the same intention. They use coarse-grained contrastive learning to put the two subsequences with the same intention closer.

Unlike the aforementioned studies, our LACLRec automatically deletes noise from the raw sequence and inserts new items to assist in learning item transition correlations, without relying on manual heuristics.

\subsection{CL-based Recommendation Framework}
Other research has proposed various CL-based recommendation frameworks to improve different recommendation tasks. \cite{40-,41-,42-,43-,44-,45-,46-} apply graph contrastive learning techniques to GNN-based recommendation and explore how to create more optimal contrastive views. \citet{47-,48-} introduce knowledge graph contrastive learning frameworks for recommendation. \citet{49-} design two CL tasks for CTR (Click-Through Rate) prediction. \cite{50-,51-,52-} propose CL-based next point-of-interest (POI) recommendation models, which aim to uncover users' latent preferences. \citet{53-} identify a vulnerability in CL-based recommendation systems, noting that they are more susceptible to poisoning attacks designed to promote specific items. \citet{54-} design a supervised contrastive learning framework to model relationships between sequences, facilitating cross-domain sequential recommendation. \cite{55-,56-,57-} propose multi-behavior contrastive learning to distill transferable knowledge from users' different types of behaviors. \citet{58-} introduce contrastive regularization to reshape the distribution of sequence representations, preventing excessive semantic similarity among item embeddings. 

Unlike the aforementioned works, we focus on the sequential recommendation task and explore finer-grained contrastive signals between augmented data and raw data, proposing a triplet contrastive learning framework.

\section{PRELIMINARIES}
\subsection{Sequential Recommender}
We denote the set of users and items as $U$ and $I$, the interaction sequence of user $u\in U$ as $S_u=\left[i_{1}, i_{2}, \ldots, i_{|S_u|}\right]$, where $i_{t}\in I$ represents the item that user interacted with at the $t$-th time step and $|S_u|$ represents the length of the sequence. The task of sequential recommender is to predict the user's next interaction item. We follow the baseline sequential recommendation framework SASRec \cite{12-kang}, mask the last position of the sequence and require the recommender to predict the item. The recommendation task is formulated as follows:
\begin{equation}
    \underset{i_{t} \in \mathcal{I}}{\arg \max } P\left(i_{\left|S_{u}\right|+1}=i_{t} \mid S_{u}\right),
\end{equation}
which selects the item $i_t$ from the candidate set $I$ that has the highest probability for recommendation.

\subsection{Random Sequence Augmentation}
In this paper, we select three random sequence augmentation methods from CL4SRec \cite{17-xie} to generate a randomly augmented sequence for each raw sequence.

\begin{itemize}[leftmargin=*]
    
    \item \textbf{Mask}. We randomly mask $\gamma |S_u|$items in the sequence according to a predefined ratio $\gamma$.
    \begin{equation}
        S_{u}^{M} = \left[i_{1}, mask, i_{3}, mask, \ldots, i_{|S_u|}\right],
    \end{equation}
    where $S_{u}^{M}$ denotes the masked sequence, $mask$ denotes the masked item.
    
    \item \textbf{Crop}. We randomly select a position in the sequence and then crop a continuous subsequence of length $\eta |S_u|$.
    \begin{equation}
        S_{u}^{C} = \left[i_{c}, i_{c+1}, \ldots, i_{c+\eta |S_u| -1}\right],
    \end{equation}
    where $i_{c}$ denotes the beginning position.

    \item \textbf{Reorder}. We randomly select a subsequence of length $\beta |S_u|$ and shuffle its items.
    \begin{equation}
        S_{u}^{R} = \left[i_{1}, i_{2}, \ldots, i_{r}^{'}, i_{r+1}^{'}, \ldots, i_{r+\beta |S_u| -1}^{'}, \ldots, i_{|S_u|}\right],
    \end{equation}
    where $\left[i_{r}^{'}, i_{r+1}^{'}, \ldots, i_{r+\beta |S_u| -1}^{'}\right]$ denotes the shuffled subsequence.
\end{itemize}

When model training, we randomly select an augmentation method for each sequence $S_u$ at each epoch, and generate an augmentation sequence.

\section{METHODS}

\begin{figure*}[h]
  \setlength{\abovecaptionskip}{0cm}
  \setlength{\belowcaptionskip}{-0.2cm}
  \centering
  \includegraphics[width=\textwidth]{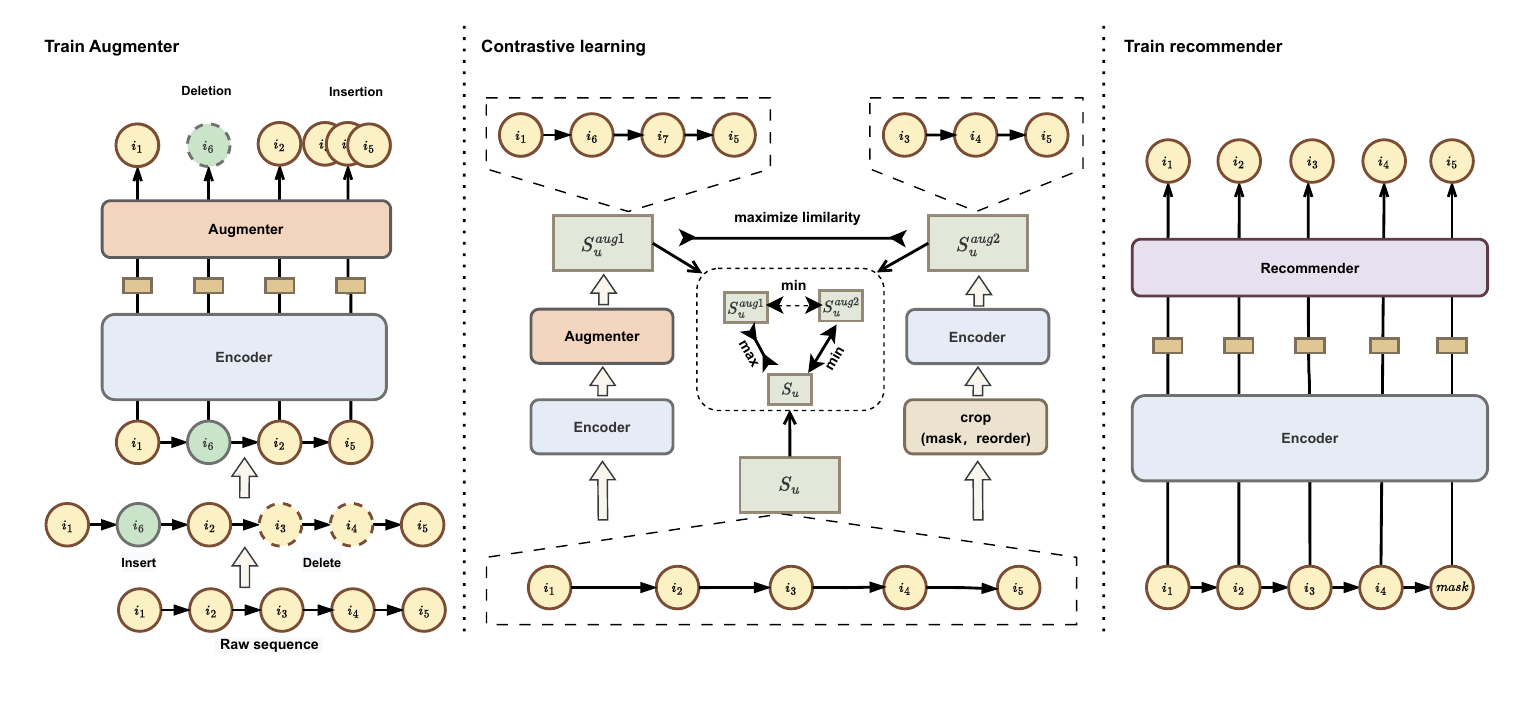}
  \caption{Overview of LACLRec. When training augmenter, we require it to restore the random modified sequences. The augmented sequence $S_u^{aug1}$ is generated by the augmenter, and $S_u^{aug2}$ is generated by random methods. Two contrastive learning tasks are used to provide supervised signals for item correlations learning. Finally, the recommender predicts the next item.}
  \Description{The figure is used to help readers understand LACLRec model.}
  \label{fig2}
\end{figure*}

\subsection{Overview}
As illustrated in Figure \ref{fig2}, LACLRec consists of four modules: a shared encoder, a self-supervised sequence augmenter, contrastive learning module, and a sequential recommender. During training, we use a Transformer-based encoder to encode the input sequence $S_u$ into embeddings, the encoder is shared between the augmenter and the recommender. The sequence augmenter then modifies the input sequence. It first determines the operation to perform on each position in $S_u$: `keep', `delete', or `insert'. For  `insert' operation, a reverse generator calculates the probability of each candidate item being inserted, and then a subsequence is inserted in reverse. This study does not consider `substitute', as this can be achieved by `delete' at the $t$-th position and `insert' at the next position. To train the augmenter, we randomly modify $S_u$ and require the augmenter to restore it through a self-supervised training task. Subsequently, for $N$ sequences in a batch, we adopt the trained augmenter to generate a SSL-augmented sequence $S_{u}^{aug1}$ for each raw sequence, and generate another augmented sequence $S_{u}^{aug2}$ by traditional random methods. In the contrastive learning module, we apply a coarse-grained contrastive loss $L_{cl}$ to increase the representation similarity between positive pairs (the two augmented sequences derived from the same raw sequence) while reducing their similarity with the other $2N-2$ negative samples. Next, we design a finer-grained triplet contrastive loss $L_{tri}$ to directly reduce the distance between the raw sequence and the SSL-augmented sequence, making the augmenter provides more self-supervised signals for recommendation compared to the random augmentation. Finally, the recommender is used to predict the user's next interaction. By jointly optimizing $L_{cl}$, $L_{tri}$, and the recommender loss $L_{rec}$, we improve the recommendation performance of LACLRec.

\subsection{Sequence Encoder}
The encoder is used to encode the input sequence into hidden representations, and it is shared by the augmenter and the recommender.

Specifically, we first define the item embedding matrix $E \in \mathbb{R}^{|I| \times e}$ to project the representation of each item into a low-dimensional dense vector, where $e$ is the dimension of the embedding vector, $I$ denotes the number of candidate items. For an item $i_t$ in the input sequence $S_u$, we index the embedding matrix $E$ to obtain its embedding vector: $e_{t} \in \mathbb{R}^{e}$, and inject position information to get its initial hidden representation $h_t^0$:
\begin{equation}
\label{eq1}
    h_{t}^{0}=e_{t}+p_{t},
\end{equation}
where $p_t$ represents the position embedding at the $t$-th position. After stacking the initial representation vectors of all items, we get the initial representation matrix $H_{e}^{0} \in \mathbb{R}^{|S_u| \times e}$.

We follow SASRec to update $H_e^0$ by a Transformer \cite{12-kang, 17-xie} with $L$ layers:
\begin{equation}
\label{eq2}
    H_{e}^{l}=\operatorname{Trm}\left(H_{e}^{l-1}\right),
\end{equation}
where $\operatorname{Trm}$ denotes a Transformer block, $H_e^l$ denotes the representation matrix at the $l$-th layer. Finally, the encoder inputs the last layer of $H_e^L$ into the augmenter and recommender. We omit the superscript $\text { (i.e., } H_{e})$ in the following sections to simplify the notation.

\subsection{SSL Sequence Augmenter}
The self-supervised sequence augmenter is used to perform data augmentation on the input sequence. Considering two issues in real-world user interaction sequences: (1) \textbf{Noisy interactions}, where items in the sequence may not reflect true item relevance (e.g., promotions or group purchases for discounts), and (2) \textbf{Data sparsity}, where many items strongly related to user preferences are not interacted with by the user, we designed three augmentation operations: `keep', `delete', and `insert'.

As shown in Figure 3, for an item $i_t$ in the sequence $S_u$, the augmenter first calculates the probability distribution for the three operations and selects the operation with the highest probability. If the chosen operation is `keep', the augmenter skips the item. If the operation is `delete', the item is deleted. If the operation is `insert', the augmenter uses a reverse generator to insert new items before $i_t$. We follow Eq. \ref{eq7} to calculate the probability distribution of performing the three operations.
\begin{equation}
\label{eq7}
    P\left(\hat{o}_{t} \mid S\right)=\operatorname{softmax}\left(W h_{t}\right),
\end{equation}
where $h_t \in {\mathbb{R}^e}$ is the representation of $i_t$, $\hat{o}_{t}$ denotes the predicted operation, $W \in \mathbb{R}^{3 \times e}$ is a projection matrix.

The augmenter applies a reverse generator to perform the insert operation. Unlike other data augmentation methods, we allow the insertion of a subsequence with a maximum length of $n$ at the selected position, rather than just a single item. The currently generated inserted sequence is denoted as $S_{1: n-1}^{<i_{t}}$. The augmenter first indexes the embedding vector of each item in $S_{1: n-1}^{<i_{t}}$, then stacks the hidden representation $h_t$ and the embedding vector of each item in $S_{1: n-1}^{<i_{t}}$, while adding the position embedding:
\begin{equation}
\label{eq8}
    H_{c}^{0}=\left[\begin{array}{c}h_{t}+p_{1} \\e_{1}+p_{2} \\\ldots \\e_{n-1}+p_{n}\end{array}\right],
\end{equation}
where $H_c^0$ denotes the initial representation matrix of reverse generator. We also adopt dropout to $H_c^0$.

We then updates $H_c^0$ by Transformer to obtain its representation $H_c \in \mathbb{R}^{n \times e}$ at the final layer, and calculate the probability distribution of the next inserted item $i_n$:
\begin{equation}
\label{eq9}
    P\left(\hat{i}_{n} \mid S_{1: n-1}^{<i_{t}}\right)=\operatorname{softmax}\left(E h_{n}\right),
\end{equation}
where $h_n \in \mathbb{R}^e$ is the hidden representation of the last position of $H_c$. In particular, the first inserted item $i_1$ is generated based on $H_c^0=\left[h_t+p_1\right]$, so the probability is expressed as $P\left(\hat{i}_{1} \mid S\right)$.

We train the augmenter using a random modification and restoration task. Specifically, 
we randomly modify the raw sequence with probabilities of keep: $p_k$, delete: $p_d$, insert: $p_i$, $p_k+p_d+p_i=1$, generating a randomly modified sequence $S_m$. For the inserted items, the augmenter needs to delete them; for the deleted items, the augmenter not only needs to perform insert operation but also accurately predict all the deleted items. This process is repeated for each raw sequence, allowing the augmenter to be trained in a self-supervised manner, learning the ability to automatically delete noise and insert strongly related new items. As shown in Eq. \ref{eq10}, the target loss function of the augmenter is to minimize the negative log-likelihood of the probability $P\left(S_{u} \mid S_{m}\right)$:
\begin{equation}
\label{eq10}
\resizebox{1.0\hsize}{!}{$
    \begin{aligned}L_{aug} & =-\log P\left(S_{u} \mid S_{m}\right) \\& =-\left(\log P\left(O \mid S_{m}\right)+\sum_{i \in I^{i n s}} \log P\left(S^{<i} \mid S_{m}\right)\right) \\& =-\left(\sum_{t=1}^{\left|S_{m}\right|} \log P\left(\hat{o}_{t}=o_{t} \mid S_{m}\right)+\sum_{i \in I^{i n s}} \sum_{n=1}^{\left|S^{<i}\right|} \sum_{j=1}^{|I|} \log P\left(\hat{i}_{j}=i_{n} \mid S_{1: n-1}^{<i}, S_{m}\right)\right)\end{aligned}$},
\end{equation}
$I^{ins}$ indicates the positions where items need to be inserted, $S^{<i}$ represents the ground truth of the inserted items.

\subsection{Triplet Contrastive Learning} 
We first apply a coarse-grained contrastive loss to provide supervisory signals for model training. The optimization objective of the contrastive loss is to maximize the similarity between two augmented sequences derived from the same user interaction sequence, while minimizing the similarity between augmented sequences from different users. Assuming there are $N$ user sequences in a batch, we obtain self-supervised augmented sequences $\left[S^{aug1}_{u_1}, S^{aug1}_{u_2}, \ldots, S^{aug1}_{u_N} \right]$ and randomly augmented sequences $\left[S^{aug2}_{u_1}, S^{aug2}_{u_2}, \ldots, S^{aug2}_{u_N} \right]$. We treat $(S^{aug1}_{u}, S^{aug2}_{u})$ as the positive pair, while the other $2N-2$ sequences serve as negative samples. The contrastive loss can then be expressed as:
\begin{equation}
\label{eq11}
\resizebox{1.0\hsize}{!}{$
    L_{cl}\left(S_{u}^{aug1}, S_{u}^{aug2}\right)=-\log \frac{\exp \left(\operatorname{sim}\left(S_{u}^{aug1}, S_{u}^{aug2}\right)\right)}{\exp \left(\operatorname{sim}\left(S_{u}^{aug1}, S_{u}^{aug2}\right)\right)+\sum_{S^{-}} \exp \left(\operatorname{sim}\left(S_{u}^{aug1}, S^{-}\right)\right)}
    $},
\end{equation}
where $S^{-}$ denotes the negative samples, and $\operatorname{sim}(\cdot)$ is calculated by the dot product of the hidden representations of the sequences.

Next, we design a finer-grained contrastive loss function. Since the self-supervised augmenter demonstrates the ability to delete noisy interactions and insert highly relevant items during training, we believe that the sequence augmented by the augmenter contributes more to modeling item transition correlations compared to the randomly augmented sequence. Therefore, a triplet contrastive loss is used to amplify the supervisory signals from the augmenter. Specifically, for a user’s triplet sequence $\left[S_u, S_u^{aug1}, S_u^{aug2}\right]$, we directly maximize the similarity $\operatorname{sim}(S_u, S_u^{aug1})$ between the raw sequence and the self-supervised augmented sequence:
\begin{equation}
\label{eq12}
    \resizebox{1.0\hsize}{!}{$
    L_{tri}\left(S_u, S_u^{aug1}, S_u^{aug2}\right)=-\log \frac{\exp \left(\operatorname{sim}\left(S_{u}, S_{u}^{aug1}\right)\right)}{\exp \left(\operatorname{sim}\left(S_{u}, S_{u}^{aug1}\right)\right) + \exp \left(\operatorname{sim}\left(S_{u}, S_{u}^{aug2}\right)\right)}
    $}.
\end{equation}

We jointly optimize $L_{cl}$ and $L_{tri}$, achieving higher similarity between the two augmented sequences from the same user at a macro level. At a micro level, the augmenter provides additional supervisory signals and directly feeds the modified results back into the hidden representation of the raw sequence.

\subsection{Recommender and Joint learning}
The recommender is used to predict the user's next interaction. In this study, we follow the baseline model CL4SRec by applying a unidirectional Transformer-based recommender.

Given the input sequence $S_u$ and its hidden representation matrix $H_e$, we again update $H_e$ by a Transformer. The initial hidden representation matrix of recommender is denoted as $H_r^0 \in \mathbb{R}^{|S_u| \times e}$, and $H_r^0=H_e$. $H_r^0$ is updated following Eq. \ref{eq13}:
\begin{equation}
\label{eq13}
    H_{r}^{l}=\operatorname{Trm}\left(H_{r}^{l-1}\right),
\end{equation}
where $H_r^l$ denotes the representation matrix at $l$-th layer. We take the representation of the last layer and denote it as $H_r$.

During training, we mask the last item $i_{|S_u|}$ in $S_u$, and the recommender predicts its probability distribution:
\begin{equation}
\label{eq14}
    P\left(\hat{i}_{t} \mid S_u\right)=\operatorname{softmax}\left(E h_{t}\right),
\end{equation}
where $E$ denotes the shared item embedding matrix, $h_t \in \mathbb{R}^e$ denotes the hidden representation of the masked item from $H_r$, $\hat{i}_{t}$ denotes the predicted item. When testing, we calculate $P\left(i_{\left|S_u\right|+1} \mid S_u\right)$ to predict the next item.
The recommender is optimized by minimizing the negative log-likelihood of the probability $P\left(\hat{i}_{t} \mid S_u\right)$:
\begin{equation}
    \begin{aligned}
        L_{rec} &=-\log P\left(\hat{i}_{t} \mid S_u\right) \\& =-\log P\left(\hat{i}_{t}= i_{|S_u|} \mid S_u \right).
    \end{aligned}
\end{equation}

In practical experiments, we first optimize the augmenter independently by minimizing $L_{aug}$. Then, the standard backpropagation algorithm is adopted to minimize the joint loss $L$ and optimize the parameters of recommender:
\begin{equation}
\label{eq22}
    L=L_{rec}+\alpha L_{cl}+\beta L_{tri},
\end{equation}
where \(\alpha\) and \(\beta\) are weight hyperparameters used to control the contribution of the contrastive losses. We will discuss their impact on model performance in the experimental section.

\section{EXPERIMENTAL SETUP}
\subsection{Research Questions}
We seek to answer the following research questions: (RQ1) Does LACLRec outperform other state-of-the-art sequential recommendation models? (RQ2) Do sequence augmenter and triplet contrast enhance recommendation accuracy? (RQ3) How robust is LACLRec in handling noisy interaction sequences? (RQ4) Is LACLRec sensitive to hyperparameter settings?

\begin{table}[h]\small%
	\centering
	\caption{Statistics of the datasets.}
	\label{tab1}
	\begin{tabular}{l|c|c|c}
		\toprule
		Datasets & Beauty & Yelp & Sports\\ 
		\midrule 
		Users & 22,362 & 22,844  & 35,597 \\
		Items & 12,101 & 16,552  & 18,357    \\  
		Records  & 194,682 & 236,999  & 294,483    \\
		Avg. length & 8.7  & 10.4 & 8.3\\
		Density & 0.07\%  & 0.06\%  & 0.05\%\\
		\bottomrule 
	\end{tabular}
\end{table}

\subsection{Datasets}
We conduct experiments on three datasets: Beauty, Yelp, and Sports. Among them, Beauty and Sports are two product review datasets crawled from Amazon \cite{59-}. Yelp is a business recommendation dataset released by Yelp.com. Due to the large size of the Yelp dataset, we only use records from 2019.

We follow \cite{30-} to preprocess the datasets. First, we filtered out users and items with fewer than 5 interaction records. Then, we sorted the interaction records of each user in chronological order to obtain the item sequences. For each item sequence, the last item is designated as the test item, the second to last item as the validation item, and the remaining items are used for training. The statistics of the preprocessed datasets are shown in Table \ref{tab1}:

\subsection{Baselines}
To verify the performance of LACLRec, we compare it with the following SOTA sequential recommendation baselines, which can be classified into three groups: (1) vanilla recommendation models; (2) contrastive learning-based models, and (3) sequence modification-based models.
\begin{itemize}
    \item Vanilla recommendation models:
    \begin{itemize}
        \item SASRec \cite{12-kang} introduces the Transformer module to model the transition relationship between items.
        \item BERT4Rec \cite{13-sun} introduces BERT into sequential recommendation, using a bidirectional Transformer and trained by masked item prediction task.
    \end{itemize}

    \item Contrastive learning-based models:
    \begin{itemize}
        \item CL4SRec \cite{17-xie} employs three random sequence augmentation methods to construct contrastive learning signals for sequential recommendation.
        \item DuoRec \cite{58-} employs an augmentation method based on dropout and a novel sampling strategy to construct contrastive self-supervised signals.
        \item CoSeRec \cite{31-} designs two SSL-based sequence augmentation methods and combines them with random methods.
        \item ICSRec \cite{39-} segments user intentions from the raw sequence and introduces intent contrastive learning.
    \end{itemize}
    
    \item Sequence modification-based models:
    \begin{itemize}
        \item STEAM \cite{30-} designs a corrector to delete misclicked items and insert missed items, aiming to improve the recommendation accuracy.
    \end{itemize}
\end{itemize}

\subsection{Evaluation Metrics and Implementation}
We adopted three widely used TOP-K metrics to evaluate the performance of all the aforementioned recommendation models: HR@$K$ (Hit Rate), MRR@$K$ (Mean Reciprocal Rank), and NDCG@$K$ (Normalized Discounted Cumulative Gain), where $K$ is set to 5, 10, or 20.

For all models, the experimental setup is determined by the original paper and the parameter tuning process. We initialize the model parameters using the Xavier method \cite{61-} and train the model using the Adam optimizer, with a learning rate set to 0.001 and an embedding size set to 64. Like \cite{13-sun, 30-}, we randomly select 99 uninteracted items as negative samples for each validation and test item to evaluate the recommendation performance. The number of heads in the Transformer is set to 1, the number of layers in the network is 1, and the dropout rate is 0.5. We limit the maximum length of the raw sequence to 50, and if the length exceeds 50, the 50 most recent records are kept. The augmenter can insert up to five consecutive items at each position of the sequence, and the maximum length of the augmented sequence is limited to 60. All experiments were performed using an RTX 4090 graphics card.

\section{EXPERIMENTAL RESULTS}
\subsection{Overall Performance}

\begin{table*}[ht]
    \setlength{\abovecaptionskip}{0.2cm}
    \setlength{\belowcaptionskip}{-0.2cm}
    \fontsize{9pt}{8pt}\selectfont
    \caption{Overall performance. The best performance and the second best performance are denoted in bold and underlined fonts respectively.}
    \label{tab2}
    \begin{tabular}{c | c | cc cccc c c | c | c}
      \toprule
      \multirow{2}{*}{Dataset} & \multirow{2}{*}{Metrics} & \multirow{2}{*}{SASRec} & \multirow{2}{*}{BERT4Rec} & \multirow{2}{*}{CL4SRec} & \multirow{2}{*}{DuoRec} & \multirow{2}{*}{CoSeRec} & \multirow{2}{*}{ICSRec} & \multirow{2}{*}{STEAM} & \multirow{2}{*}{LACLRec} & \multicolumn{2}{c}{improve v.s.} \\ \cline{11-12} 
      & & & & & & & & & & \multicolumn{1}{c|}{CL4SRec} & All \\ \hline
      \midrule
      \multirow{10}{*}{Beauty} &HR5 &0.3721 &0.3666 &0.4067 &0.4094 &0.4124 &0.4250 &\underline{0.4284} &\textbf{0.4446} &9.32\% &3.78\% \\
      &HR10 &0.4639 &0.4728 &0.5056 &0.5078 &0.5187 &0.5188 &\underline{0.5256} &\textbf{0.5460} &7.99\% &3.88\% \\
      &HR20 &0.5804 &0.6011 &0.6199 &0.6200 &0.6356 &0.6345 &\underline{0.6486} &\textbf{0.6606} &6.57\% &1.85\% \\
      &MRR5 &0.2611 &0.2337 &0.2785 &0.2884 &0.2749 &\underline{0.2963} &0.2899 &\textbf{0.3161} & 13.5\% &6.68\% \\
      &MRR10 &0.2732 &0.2478 &0.2916 &0.3015 &0.2890 &\underline{0.3087} &0.3028 &\textbf{0.3297} &13.07\% &6.8\% \\
      &MRR20 &0.2812 &0.2566 &0.2994 &0.3092 &0.2971 &\underline{0.3167} &0.3112 &\textbf{0.3376} &12.76\% &6.6\% \\
      &NDCG5 &0.2887 &0.2570 &0.3104 &0.3194 &0.3090 &\underline{0.3284} &0.3181 &\textbf{0.3482} &12.18\% &6.03\% \\
      &NDCG10 &0.3182 &0.2907 &0.3422 &0.3494 &0.3434 &\underline{0.3586} &0.3508 &\textbf{0.3810} &11.34\% &6.25\% \\
      &NDCG20 &0.3476 &0.3227 &0.3710 &0.3774 &0.3728 &\underline{0.3878} &0.3807 &\textbf{0.4099} &10.49\% &5.7\% \\
      &Sum &3.1868 &3.049 &3.4253 &3.4825 &3.4529 &\underline{3.5748} &3.5561 &\textbf{3.7740} &10.18\% &5.57\% \\
      \midrule
      \multirow{10}{*}{Yelp} &HR5 &0.5847 &0.6118 &0.6292 &0.6400 &0.6454 &0.6618 &\underline{0.6683} &\textbf{0.6833} &8.6\% &2.24\% \\
      &HR10 &0.7833 &0.7972 &0.8223 &0.8263 &0.8241 &0.8337 &\underline{0.8460} &\textbf{0.8569} &4.21\% &1.17\% \\
      &HR20 &0.9194 &0.9235 &0.9486 &0.9564 &0.9444 &\underline{0.9531} &0.9511 &\textbf{0.9554} &0.72\% &0.24\% \\
      &MRR5 &0.3543 &0.3764 &0.3991 &0.4085 &0.4140 &\underline{0.4302} &0.4293 &\textbf{0.4482} &12.3\% &4.18\% \\
      &MRR10 &0.3808 &0.4013 &0.4251 &0.4334 &0.4380 &\underline{0.4533} &0.4531 &\textbf{0.4716} &10.94\% &4.04\% \\
      &MRR20 &0.3907 &0.4104 &0.4343 &0.4429 &0.4467 &\underline{0.4619} &0.4607 &\textbf{0.4788} &10.25\% &3.66\% \\
      &NDCG5 &0.4113 &0.4257 &0.4662 &0.4647 &0.4714 &0.4877 &\underline{0.4895} &\textbf{0.5066} &11.05\% &3.49\% \\
      &NDCG10 &0.4756 &0.4870 &0.5188 &0.5245 &0.5293 &0.5434 &\underline{0.5461} &\textbf{0.5630} &8.52\% &3.09\% \\
      &NDCG20 &0.5105 &0.5218 &0.5513 &0.5590 &0.5602 &\underline{0.5741} &0.5735 &\textbf{0.5883} &6.71\% &2.47\% \\
      &Sum &4.8109 &4.9551 &5.1849 &5.2557 &5.2734 &5.3993 &\underline{5.4176} &\textbf{5.5523} &7.09\% &2.49\% \\
      \midrule
      \multirow{10}{*}{Sports} &HR5 &0.3450 &0.3515 &0.3935 &0.3980 &0.4074 &\underline{0.4220} &0.4190 &\textbf{0.4379} &11.28\% &3.77\% \\
      &HR10 &0.4597 &0.4791 &0.5152 &0.5192 &0.5305 &0.5459 &\underline{0.5496} &\textbf{0.5645} &9.57\% &2.71\% \\
      &HR20 &0.5942 &0.6280 &0.6562 &0.6583 &0.6650 &0.6861 &\underline{0.6916} &\textbf{0.7017} &6.93\% &1.46\% \\
      &MRR5 &0.2204 &0.2154 &0.2600 &0.2597 &0.2653 &\underline{0.2741} &0.2685 &\textbf{0.2896} &11.38\% &5.65\% \\
      &MRR10 &0.2357 &0.2323 &0.2762 &0.2758 &0.2817 &\underline{0.2906} &0.2858 &\textbf{0.3064} &10.93\% &5.44\% \\
      &MRR20 &0.2449 &0.2426 &0.2859 &0.2854 &0.2910 &\underline{0.3003} &0.2956 &\textbf{0.3159} &10.49\% &5.19\% \\
      &NDCG5 &0.2513 &0.2448 &0.2931 &0.2936 &0.3005 &\underline{0.3108} &0.3058 &\textbf{0.3264} &11.36\% &5.02\% \\
      &NDCG10 &0.2883 &0.2862 &0.3324 &0.3335 &0.3403 &\underline{0.3508} &0.3479 &\textbf{0.3673} &10.5\% &4.7\% \\
      &NDCG20 &0.3222 &0.3236 &0.3680 &0.3686 &0.3743 &\underline{0.3862} &0.3838 &\textbf{0.4020} &9.24\% &4.09\% \\
      &Sum &2.9622 &3.0035 &0.3805& 3.3921 &3.4559 &\underline{3.5668} &3.5476 &\textbf{3.7122} &9.81\% &4.08\% \\
      \bottomrule
    \end{tabular}
\end{table*}

To answer RQ1, we first compare the overall performance of LACLRec with seven baseline models, the experimental results are shown in Table \ref{tab2}. From the experimental results, we have the following observations. 

First, LACLRec achieves the best scores across all nine evaluation metrics on the three datasets, demonstrating its superior recommendation performance. Specifically, the sum of metrics improves by up to 10.18\% compared to the baseline contrastive learning model CL4SRec, and by up to 5.57\% compared to the second-best model. Unlike other models, LACLRec utilizes a learnable sequence augmenter to perform insertion and deletion operations, generating new augmented sequences for the contrastive learning task. On one hand, the sequence augmenter provides a strong contrastive signal, with the generated augmented sequences alleviating issues of data sparsity and noisy interactions in the raw sequence. On the other hand, the triplet contrastive learning provides finer-grained self-supervision signals, enhancing the recommender's ability to capture item representations and item correlations.

Second, the contrastive learning models outperform the vanilla models on all datasets, with ICSRec achieving the second-best performance on most metrics. This indicates that contrastive learning tasks can extract additional supervisory signals from user interaction sequences, enhancing the performance of sequential recommendation. LACLRec surpasses these models by employing a learnable augmenter rather than random or manual methods, generating higher-quality augmented sequences whose effectiveness is further amplified through triplet contrast.

Lastly, STEAM introduces the `non-exposed' and `misclicks' in real user interactions \cite{30-}, designing a corrector to automatically delete noisy interactions and insert new items. In comparison, STEAM lacks the stronger supervisory signals provided by contrastive learning, resulting in suboptimal performance comparable to ICSRec.

\subsection{Ablation Study}

\begin{table*}[ht]
    \setlength{\abovecaptionskip}{0.2cm}
    \setlength{\belowcaptionskip}{-0.2cm}
    \fontsize{9pt}{8pt}\selectfont
    \caption{Ablation Study. The best performance is denoted in bold fonts.}
    \label{tab3}
    \begin{tabular}{c|c| ccc ccc ccc c}
         \toprule
         Dataset & Model & HR5 & HR10 & HR20 & MRR5 & MRR10 & MRR20 & NDCG5 & NDCG10 & NDCG20 & Sum \\
         \midrule
         \multirow{6}{*}{Beauty} & LACLRec & \textbf{0.4446} &\textbf{0.5460} &\textbf{0.6606} &\textbf{0.3161} &\textbf{0.3297} &\textbf{0.3376} &\textbf{0.3482} &\textbf{0.3810} &\textbf{0.4099} &\textbf{3.7740} \\
         &Base & 0.4066 &0.5067 &0.6202 &0.2806 &0.2939 &0.3017 &0.3119 &0.3442 &0.3729 &3.4391 \\
         &$w/o-Tri$ & 0.4311 &0.5263 &0.6429 &0.3076 &0.3203 &0.3283 &0.3384 &0.3691 &0.3985 &3.6628 \\
         &DuoAug &0.4097 &0.5032 &0.6173 &0.2943 &0.3067 &0.3145 &0.3230 &0.3531 &0.3818 &3.5041 \\
         &TestAug &0.4297 &0.5319 &0.6504 &0.2949 &0.3085 &0.3167 &0.3284 &0.3615 &0.3914 &3.6137 \\
         &Co-Train &0.4381 &0.5332 &0.6529 &0.3102 &0.3229 &0.3311 &0.3421 &0.3728 &0.4029 &3.7066 \\
         \midrule
         \multirow{6}{*}{Yelp} & LACLRec & \textbf{0.6833} &\textbf{0.8569} &\textbf{0.9554} &\textbf{0.4482} &\textbf{0.4716} &\textbf{0.4788} &\textbf{0.5066} &\textbf{0.5630} &\textbf{0.5883} &\textbf{5.5523} \\
         &Base & 0.6319 &0.8216 &0.9482 &0.4012 &0.4267 &0.4360 &0.4584 &0.5199 &0.5525 &5.1968 \\
         &$w/o-Tri$ & 0.6661 &0.8378 &0.9436 &0.4355 &0.4586 &0.4663 &0.4927 &0.5485 &0.5757 &5.4252 \\
         &DuoAug & 0.6354 &0.8138 &0.9339 &0.4051 &0.4290 &0.4377 &0.4622 &0.5199 &0.5508 &5.1881 \\
         &TestAug &0.6649 &0.8396 &0.9443 &0.4308 &0.4543 &0.4620 &0.4889 &0.5456 &0.5726 &5.4035 \\
         &Co-Train &0.6744 &0.8481 &0.9548 &0.4379 &0.4612 &0.4690 &0.4966 &0.5529 &0.5803 &5.4755 \\
         \midrule
         \multirow{6}{*}{Sports} & LACLRec &\textbf{0.4379} &\textbf{0.5645} &\textbf{0.7017} &\textbf{0.2896} &\textbf{0.3064} &\textbf{0.3159} &\textbf{0.3264} &\textbf{0.3673} &\textbf{0.4020} &\textbf{3.7122} \\
         &Base & 0.3946 &0.5161 &0.6585 &0.2615 &0.2776 &0.2875 &0.2945 &0.3337 &0.3697 &3.3941 \\
         &$w/o-Tri$ &0.4157 &0.5403 &0.6751 &0.2766 &0.2932 &0.3025 &0.3112 &0.3514 &0.3854 &3.5519 \\
         &DuoAug & 0.3897 &0.5119 &0.6504 &0.2578 &0.2739 &0.2835 &0.2905 &0.3298 &0.3648 &3.3528 \\
         &TestAug &0.4182 &0.5470 &0.6860 &0.2719 &0.2890 &0.2986 &0.3083 &0.3498 &0.3849 &3.5542 \\
         &Co-Train &0.4352 &0.5588 &0.6980 &0.2843 &0.3006 &0.3102 &0.3217 &0.3615 &0.3966 &3.6674 \\
         \bottomrule
    \end{tabular}
\end{table*}

To answer RQ2, we design several variant models and compared their performance. First, to validate the effectiveness of the augmenter and triplet contrast, we design two variant models:
\begin{itemize}
    \item Base: The augmenter is removed, and both augmented sequences are generated randomly.
    \item $w/o-Tri$: The augmenter is used to generate one augmented sequence, but triplet contrast is removed.
\end{itemize}
Next, we design three variant models to explore the optimal module combinations and training methods:
\begin{itemize}
    \item DuoAug: We add random perturbations in the augmenter, and use it to generate both two augmented sequences.
    \item TestAug: When testing, the recommender directly predicts the next item based on augmented sequences generated by the augmenter.
    \item Co-Train: The augmenter and recommender are jointly trained.
\end{itemize}

The experimental results are shown in Table \ref{tab3}, from which we have the following observations. Firstly, without the self-supervised augmenter, the Base actually degrades into CL4SRec, with a significant decrease in recommendation performance. The performance of $w/o-Tri$ also noticeably declines compared to LACLRec, indicating that triplet contrastive learning provides finer-grained supervisory signals that benefit next-item prediction tasks. Secondly, DuoAug performs poorly, due to the high similarity between the two augmented sequences generated by the augmenter from the same raw sequence, which fails to provide effective contrastive signals. TestAug predicts items based on augmented sequences, also shows inferior performance since the augmented sequences do not directly participate in recommender training. If the next item is predicted based on the augmented sequence, it will reduce the efficiency of model training and inference \cite{30-}. Finally, co-training the augmenter and recommender also decreases performance. We attribute this to the early stages of training, where the untrained augmenter may generate low-quality augmented sequences, thereby misleading the contrastive learning task.

In conclusion, both the augmenter and triplet contrastive learning effectively improve the accuracy of sequential recommendations, and optimizing the model is best achieved by employing both modules while training the augmenter and recommender separately.

\subsection{Robustness analysis}

\begin{table*}[ht]
    \setlength{\abovecaptionskip}{0.2cm}
    \setlength{\belowcaptionskip}{-0.2cm}
    \fontsize{9pt}{8pt}\selectfont
    \caption{Robustness analysis. The table shows the performance comparison of different models on the simulated test sets.}
    \label{tab4}
    \begin{tabular}{c|c| ccc ccc ccc |c|c|c}
         \toprule
         Dataset & Model & HR5 & HR10 & HR20 & MRR5 & MRR10 & MRR20 & NDCG5 & NDCG10 & NDCG20 & Sum & Raw & dist \\
         \hline
         \midrule
         \multirow{6}{*}{Beauty} & LACLRec &\textbf{0.4306} &\textbf{0.5276} &\textbf{0.6424} &\textbf{0.3057} &\textbf{0.3186} &\textbf{0.3265} &\textbf{0.3368} &\textbf{0.3681} &\textbf{0.3971} &\textbf{3.6540} &\textbf{3.7740} &\textbf{-3.17\%} \\
         &ICSRec &0.4045 &0.5100 &0.6278 &0.2709 &0.2849 &0.2931 &0.3041 &0.3381 &0.3679 &3.4013 &3.5748 &-4.85\% \\
         &STEAM &0.4081 &0.5080 &0.6279 &0.2718 &0.2851 &0.2933 &0.3057 &0.3379 &0.3681 &3.4062 &3.5561 &-4.21\% \\
         &CoSeRec &0.3925 &0.5020 &0.6262 &0.2567 &0.2714 &0.2799 &0.2905 &0.3259 &0.3572 &3.3022 &3.4529 &-4.36\% \\
         &CL4SRec &0.3911 &0.4882 &0.6008 &0.2690 &0.2819 &0.2896 &.2993 &0.3307 &0.3590 &3.3100 &3.4253 &-3.36\% \\
         &DuoRec &0.3925 &0.4928 &0.6040 &0.2705 &0.2839 &0.2915 &0.3009 &0.3333 &0.3613 &3.3311 &3.4825 &-4.34\% \\ 
         \midrule
         \multirow{6}{*}{Yelp} &LACLRec &\textbf{0.6536} &\textbf{0.8357} &0.9514 &\textbf{0.4193} &\textbf{0.4437} &\textbf{0.4522} &\textbf{0.4774} &\textbf{0.5364} &\textbf{0.5662} &\textbf{5.3365} &\textbf{5.5523} &\textbf{-3.88\%} \\
         &ICSRec &0.6183 &0.8092 &\textbf{0.9551} &0.3875 &0.4129 &0.4235 &0.4447 &0.5064 &0.5438 &5.1014 &5.3993 &-5.51\% \\
         &STEAM &0.6328 &0.8160 &0.9355 &0.3991 &0.4237 &0.4324 &0.4571 &0.5165 &0.5472 &5.1606 &5.4176 &-4.74\% \\
         &CoSeRec &0.6148 &0.7966 &0.9293 &0.3853 &0.4097 &0.4192 &0.4421 &0.5011 &0.5350 &5.0331 &5.2734 &-4.56\% \\
         &CL4SRec &0.6003 &0.7910 &0.9280 &0.3721 &0.3977 &0.4077 &0.4286 &0.4905 &0.5257 &4.9419 &5.1849 &-4.68\% \\
         &DuoRec &0.6063 &0.8011 &0.9525 &0.3757 &0.4017 &0.4128 &0.4329 &0.4959 &0.5348 &5.0141 &5.2557 &-4.59\% \\ 
         \midrule
         \multirow{6}{*}{Sports} &LACLRec &\textbf{0.4186} &\textbf{0.5440} &\textbf{0.6803} &\textbf{0.2740} &\textbf{0.2906} &\textbf{0.3000} &\textbf{0.3099} &\textbf{0.3503} &\textbf{0.3847} &\textbf{3.5527} &\textbf{3.7122} &\textbf{-4.29\%} \\
         &ICSRec &0.3988 &0.5246 &0.6701 &0.2552 &.2719 &0.2820 &0.2908 &0.3314 &0.3682 &3.3931 &3.5668 &-4.87\% \\
         &STEAM &0.3929 &0.5259 &0.6724 &0.2475 &0.2652 &0.2753 &0.2836 &0.3265 &0.3635 &3.3530 &3.5476 &-5.48\% \\
         &CoSeRec &0.3813 &0.5019 &0.6377 &0.2420 &0.2580 &0.2674 &0.2765 &0.3154 &0.3497 &3.2300 &3.4559 &-6.53\% \\
         &CL4SRec &0.3648 &0.4854 &0.6251 &0.2404 &0.2563 &0.2659 &0.2712 &0.3100 &0.3452 &3.1647 &3.3805 &-6.38\% \\
         &DuoRec &0.3746 &0.4961 &0.6362 &0.2398 &0.2558 &0.2655 &0.2732 &0.3123 &0.3477 &3.2016 &3.3921 &-5.62\% \\ 
         \bottomrule
    \end{tabular}
\end{table*}

Real-world user interaction records are often subject to noise interference due to privacy and other concerns, which may degrade the performance of trained recommendation models. To answer RQ3, we conduct several experiments on three simulated test sets to assess the robustness of LACLRec. Specifically, we apply   `keep', `delete', and `insert' operations in a 4:3:3 ratio to each item in the real test set (keeping the final item unchanged) to generate the simulated test sets. We then compare the recommendation performance of LACLRec against several strong baselines on these simulated test sets, which contain substantial noise. 

The experimental results are presented in Table \ref{tab4}, where Sum denotes the total score of the evaluation metrics on the simulated test set, and Raw denotes the total score on the real test set. First, we find that all models have noticeable performance degradation on the simulated test set due to the massive destruction of data. Nonetheless, LACLRec continues to perform best on all three datasets, with only a slight drop in the HR20 metric on the Yelp dataset compared to ICSRec, further proving the superior performance of LACLRec. Additionally, to further quantify the impact of noise on each model, we also visually show the disturbance of their total score on the simulated test set compared with the total score on the real test set, where $dist=(Sum-Raw)/Raw$. It is evident that LACLRec is the most robust, with the smallest decrease in total score across all three datasets. We attribute this robustness to the augmenter, which generates augmented sequences with higher diversity compared to traditional random augmentation methods, thereby conveying richer item correlations to the recommender. Consequently, LACLRec exhibits reduced sensitivity to noisy interactions. ICSRec, CoSeRec, and DuoRec also adopt optimized contrastive signal construction methods, which perform better overall than CL4SRec. Finally, STEAM corrects the input sequence for the recommender, which can effectively mitigate noise to some extent and showcase competitive performance ae well.

\begin{table*}[ht]
    \setlength{\abovecaptionskip}{0.2cm}
    \setlength{\belowcaptionskip}{-0.2cm}
    \fontsize{9pt}{8pt}\selectfont
    \caption{Hyperparameter Study. The table shows the effect of $[p_k, p_d, p_i]$ when training the augmenter.}
    \label{tab5}
    \begin{tabular}{c|c| ccc ccc ccc |c|c}
         \toprule
         Dataset & $[p_k, p_d, p_i]$ & HR5 & HR10 & HR20 & MRR5 & MRR10 & MRR20 & NDCG5 & NDCG10 & NDCG20 & Sum & Operation \\
         \hline
         \midrule
         \multirow{3}{*}{Beauty} &[0.4, 0.5, 0.1] &0.4446 &0.5460 &0.6606 &0.3161 &0.3297 &0.3376 &0.3482 &0.3810 &0.4099 &3.7740 &[53\%, 1\%, 46\%] \\
         &[0.5, 0.4, 0.1] &0.4439 &0.5470 &0.6631 &0.3137 &0.3274 &0.3354 &0.3461 &0.3795 &0.4087 &3.7652 &[58\%, 1\%, 41\%] \\
         &[0.5, 0.3, 0.2] &0.4431 &0.5450 &0.6571 &0.3147 &0.3282 &0.3359 &0.3467 &0.3796 &0.4078 &3.7587 &[74\%, 1\%, 25\%] \\
         \midrule
         \multirow{3}{*}{Yelp} &[0.4, 0.5, 0.1] &0.6833 &0.8569 &0.9554 &0.4482 &0.4716 &0.4788 &0.5066 &0.5630 &0.5883 &5.5523 &[50\%, 2\%, 48\%] \\
         &[0.5, 0.4, 0.1] &0.6781 &0.8505 &0.9556 &0.4444 &0.4676 &0.4753 &0.5024 &0.5584 &0.5855 &5.5180 &[72\%, 1\%, 27\%] \\
         &[0.5, 0.3, 0.2] &0.6767 &0.8511 &0.9578 &0.4406 &0.4641 &0.4719 &0.4992 &0.5558 &0.5833 &5.5008 &[83\%, 4\%, 13\%] \\
         \midrule
         \multirow{3}{*}{Sports} &[0.4, 0.5, 0.1] &0.4336 &0.5626 &0.7016 &0.2877 &0.3049 &0.3145 &0.3240 &0.3656 &0.4007 &3.6956 &[49\%, 1\%, 50\%] \\
         &[0.5, 0.4, 0.1] &0.4379 &0.5645 &0.7017 &0.2896 &0.3064 &0.3159 &0.3264 &0.3673 &0.4020 &3.7122 &[61\%, 1\%, 38\%] \\
         &[0.5, 0.3, 0.2] &0.4355 &0.5628 &0.7005 &0.2883 &0.3052 &0.3148 &0.3248 &0.3660 &0.4008 &3.6992 &[83\%, 2\%, 15\%] \\
         \bottomrule
    \end{tabular}
\end{table*}

\begin{table*}[ht]
    \setlength{\abovecaptionskip}{0.2cm}
    \setlength{\belowcaptionskip}{-0.2cm}
    \fontsize{9pt}{8pt}\selectfont
    \caption{Hyperparameter Study. The table shows the effect of $\alpha$ and $\beta$ when training the recommender.}
    \label{tab6}
    \begin{tabular}{c|c| ccc ccc ccc |c}
         \toprule
         Dataset & $\alpha, \beta$ & HR5 & HR10 & HR20 & MRR5 & MRR10 & MRR20 & NDCG5 & NDCG10 & NDCG20 & Sum \\
         \hline
         \midrule
         \multirow{7}{*}{Beauty} &0.1, 0.005 &0.4446 &0.5460 &0.6606 &0.3161 &0.3297 &0.3376 &0.3482 &0.3810 &0.4099 &3.7740 \\
         &0.1, 0.01 &0.4466 &0.5491 &0.6608 &0.3147 &0.3283 &0.3359 &0.3476 &0.3806 &0.4087 &3.7727 \\
         &0.1, 0.05 &0.4352 &0.5349 &0.6495 &0.3048 &0.3180 &0.3259 &0.3373 &0.3694 &0.3984 &3.6738 \\
         &0.1, 0.1 &0.4181 &0.5228 &0.6340 &0.2834 &0.2973 &0.3050 &0.3170 &0.3507 &0.3788 &3.5076 \\
         &0.2, 0.005 &0.4418 &0.5426 &0.6551 &0.3124 &0.3258 &0.3336 &0.3446 &0.3772 &0.4056 &3.7390 \\
         &0.5, 0.005 &0.4410 &0.5342 &0.6477 &0.3111 &0.3235 &0.3313 &0.3435 &0.3735 &0.4022 &3.7084 \\
         &1.0, 0.005 &0.4395 &0.5349 &0.6463 &0.3112 &0.3239 &0.3315 &0.3431 &0.3739 &0.4020 &3.7066 \\
         \bottomrule
    \end{tabular}
\end{table*}

\subsection{Hyperparameter Study}
LACLRec has two unique and critical hyperparameters: the proportion of random sequence modifications during augmenter training, and the weights of the contrastive loss during recommender training. To answer RQ4, we investigate whether LACLRec is sensitive to the settings of these hyperparameters. 

First, we select three sets of $[p_k, p_d, p_i]$, with results shown in Table \ref{tab5}, where `Operation' represents the proportion of keep, delete, and insert operations when the trained augmenter generates augmented sequences. We observe that varying the random modification ratios can influence the augmenter’s modification tendencies in practical applications. Specifically, increasing the rate of random deletions encourages the augmenter to insert items more frequently. From the results, we see that a higher insertion ratio can slightly improve LACLRec's performance, likely due to augmented sequences encompassing more diverse item transitions. However, LACLRec's recommendation performance does not display significant changes with variations in the `Operation' ratio, indicating that the model is insensitive to the settings of $[p_k, p_d, p_i]$.

Next, we examine the impact of the contrastive loss weights $\alpha$ and $\beta$ on model performance, with results shown in Table \ref{tab6}. We observe similar performance variations on the three datasets, and present only the results on Beauty here. We find that as $\alpha$ and $\beta$ increase, overall recommendation performance gradually declines. The model is particularly sensitive to the triplet contrastive loss weight $\beta$. When $\beta$ increases to 0.05 or 0.1, LACLRec underperforms compared to some baseline models in Table \ref{tab2}. This is due to the fact that the scale of $L_{tri}$ is larger than that of $L_{rec}$ in our experiments, and jointly optimizing them requires a smaller weight to balance $L_{tri}$. This sensitivity to $\beta$ is a limitation of LACLRec, necessitating careful selection of the weight for $L_{tri}$.

\section{CONCLUSION}
This paper proposes LACLRec, which employs a learnable augmenter in place of traditional random methods to generate higher-quality augmented sequences for contrastive learning tasks, thereby mitigating issues of data sparsity and noisy interactions. Additionally, we design a triplet contrastive loss to provide finer-grained supervision signals for model training. Experiments on three public datasets demonstrate that LACLRec outperforms state-of-the-art baseline models in recommendation performance. Limitations of our work include the need to pre-collect data to train the augmenter and the additional hyperparameter tuning required.

\begin{acks}
This research was supported by
\end{acks}

\bibliographystyle{ACM-Reference-Format}
\bibliography{sample-base.bib}


\end{document}